# The role of charge traps in inducing hysteresis: capacitance – voltage measurements on top gated bilayer graphene

Gopinadhan Kalon, Young Jun Shin, Viet Giang Truong, Alan Kalitsov, and Hyunsoo Yang [a]

*Department of Electrical and Computer Engineering, NUSNNI-Nanocore, National University of Singapore, 117576 Singapore*

Understanding the origin of hysteresis in the channel resistance from top gated graphene transistors is important for transistor applications. Capacitance - voltage measurements across the gate oxide on top gated bilayer graphene show hysteresis with a charging and discharging time constant of ~100 µs. However, the measured capacitance across the graphene channel does not show any hysteresis, but shows an abrupt jump at a high channel voltage due to the emergence of an order, indicating that the origin of hysteresis between gate and source is due to charge traps present in the gate oxide and graphene interface.

[a] e-mail address: eleyang@nus.edu.sg

Graphene, a two dimensional material system, consists of either a single or many layers of carbon atoms arranged in a honey comb lattice. Electronically graphene is found to be a zero gap semiconductor with its conduction and valence bands touch each other at the neutrality point making it ambipolar.[1, 2] Graphene has many unique physical properties such as a very high mobility and conductivity, which makes it the technologically most sought material after silicon. However, a large surface to volume ratio allows graphene to attract many adsorbates very easily which influence the electronic properties dramatically.[3] Field effect studies of graphene on $SiO_2$ show a hysteresis in the conductance and the hysteresis varies depending upon the sweeping voltage range, sweep rate, and the surrounding conditions.[4-6] Its origin is attributed to the adsorption of $H_2O$ molecules, charge injection into the trap sites of $SiO_2$ substrate, etc.[5] Methods such as thermal annealing, vacuum treatment, and treatment of $SiO_2$ with a hydrophobic polymer are suggested to reduce the hysteresis.[7]

However, there are only a few reports on the hysteresis of the top gated bilayer graphene field effect transistor (GFET).[5] It is technologically important to understand the origin of hysteresis in top gated graphene transistors as this will serve a platform for the characterization of high frequency GFET. In all the previous studies, either resistance or conductance as a function of gate voltage is used to understand the origin of hysteresis and there is no report on the hysteresis in capacitance-voltage (C-V) measurements, especially in bilayer graphene. Quantum capacitance, a property of low dimensional systems, is proportional to the density of states and the capacitance measurements are very useful in detecting localized states of disordered systems, whose contribution to conductivity is suppressed. Capacitance measurements are reported by many groups on single layer graphene[8-10] whereas very few reports exist on bilayer graphene.[11, 12] In the case of bilayer graphene, reported C-V results are contradicting each other with one report[12] showed a V-shaped capacitance with gate voltage, whereas the other report[11] showed an inverted V-shape of capacitance with gate voltage. In this letter, we report the hysteresis by C-V measurements on top gated bilayer graphene which provide a direct experimental evidence of the existence of charge traps as the cause for the hysteresis. We also report an abrupt jump in capacitance at high applied channel voltage due to an increase in order by removal of defects in the graphene channel.

The graphene is obtained by micromechanical exfoliation of Kish graphite subsequently transferring it to a highly p-doped Si substrate, which has a layer of 300 nm thick $SiO_2$. The resultant graphene is identified by an optical microscope and confirmed by Raman spectrophotometer.[13, 14] Electrodes are prepared by optical lithography followed by the deposition of Cr (5 nm)/Au (150 nm) using a thermal evaporator. Standard lift-off procedures are followed after the deposition [see Fig. 1(a)]. For the top gate fabrication, similar procedures are followed and a 10 nm of Al is deposited in two steps followed by natural oxidation. Figure 1(b) shows the Raman spectrum of pristine graphene. The spectrum shows prominently the G peak along with a 2D peak. A Lorentian fit of a 2D peak in the inset of Fig. 1(b) estimates the full width at half maximum as 50 $cm^{-1}$, indicating bilayer graphene. The sample has no detectable D peak suggesting the absence of microscopic disorder in the graphene channel.

To understand the electron transport properties, measurements are carried out in two point geometry and under high vacuum ($< 1\times10^{-7}$ Torr) conditions. Figure 1(c) shows the channel resistance $R_{xx}$ between source (S) and drain (D) as a function of top gate (G) voltage, $V_{TG}$ at 300 K. The top gate voltage is continuously varied from 0 to 1 V, then to -1 V and finally back to 0 V. The bias voltage range is extended to ±2 and ±3 V with the same sweeping sequence. Each hysteresis loop is repeated twice to confirm the reproducibility of the results.

Two important points are noticeable in the figure; (i) charge neutrality point (CNP) corresponding to maximum resistance in the $R_{xx}$ vs. $V_{TG}$ graph is not at zero, but depends on the extent of the applied $V_{TG}$, (ii) the area under the hysteresis is a function of the extent of the $V_{TG}$. When a positive $V_{TG}$ is applied, the accumulation of electrons in charge traps causes the maximum of $R_{xx}$ to be on a positive bias voltage. With negative $V_{TG}$ due to the injection of holes into charge traps, CNP is shifted towards more negative voltage. The larger the applied positive (negative) voltage, the more the accumulation of electrons (holes), which lead the CNP to depend on the magnitude of the applied voltage. The occupancy of the traps depends on the maximum applied voltage which in turn determines the shift in CNP, whereas the area under the hysteresis is determined by the number of traps charged on applying a definite voltage. The trap density is calculated from the shift of the CNP ($\Delta V_{NP}$) from zero gate voltage using the relationship, $n_{it}=C_{ox}\Delta V_{NP}/2e$, where $C_{ox}$ is the geometric capacitance per unit area (for $Al_2O_3$, the dielectric constant is 8 and the thickness is 10 nm) and $e$ is the charge of an electron. For a maximum gate voltage of 3 V, $n_{it}$ is estimated to be $\sim 5\times10^{11}$ cm$^{-2}$ which is in agreement with the reported results on the magnitude of the charge density inhomogeneity ($2-15\times10^{11}$ cm$^{-2}$) in single and bilayer graphene.[15-17] To further understand the origin of hysteresis, capacitance measurements are performed. Figure 1(d) shows the measured capacitance (C) with top gate as one electrode and source as the ground electrode as a function of $V_{TG}$ at 300 K. The capacitance behavior is similar to the reciprocal of $R_{xx}$ and is also a function of the extent of the applied $V_{TG}$. The identical nature of C and $1/R_{xx}$ indicates that the appearance of hysteresis and existence of different CNPs is due to a capacitive source and we find that, as we discuss later, charge traps present at the graphene/gate oxide interface are causing these effects.

The different contributions to the total capacitance C is decomposed into the top gate oxide capacitance ($C_{ox}$), in series with the quantum capacitance of graphene ($C_Q$) and the trap capacitance ($C_{tr}$). It is known that $C_{ox}$ is constant and independent of the applied gate voltages, whereas the graphene quantum capacitance $C_Q$ is a measure of the response of the charges inside the channel to the change in the density of states of conduction and valence bands. Figure 2(a) shows the two probe resistance $R_{xx}$ as a function of $V_{TG}$ at 3.8 K, indicating that the hysteresis is still present at 3.8 K. A hysteresis was observed in graphene by other groups at ambient conditions, but it was suppressed either by introducing vacuum or cryogenic temperatures suggesting the removal of attached molecules in graphene.[4-7] In our samples, the hysteresis still exists even after exposing to the above conditions implying that chemical attachment is not a dominant source of the hysteresis. Figure 2(b) shows the measured C as a function of $V_{TG}$ at 3.8 K, showing a similar behavior with $1/R_{xx}$ as observed at 300 K. Figure 2(c) shows the total C between gate and source as a function of frequency $f$ and $V_{TG}$ at 300 K. The capacitance shows a sharp reduction when $f$ increases from 1 kHz to 20 kHz, but the values remain almost constant for $f$ > 20 kHz. A sharp fall at f ~ 10 kHz suggests that the charging and discharging time for the interface trap on graphene/$Al_2O_3$ is ≥ 100 μs. A recent study of graphene on $SiO_2$ estimates trapping time constants of 87 μs - 1.6 ms which is in good agreement with our result.[18] It is interesting to note that the trap time constant in graphene/oxide is much larger than that of Si/gate oxide (< 1μs).[19] Figure 2(d) shows the total capacitance between gate and source as a function of $V_{TG}$ at 100 kHz and 1 MHz. The capacitance does not follow $V_{TG}$ in contrast to that of low frequencies, which implies that electron trapping cannot follow the speed of band movements at high frequencies. This result suggests that the trapping speed can be very different depending on the type of the carriers (electrons or holes).

Figure 3(a) shows C measured at 10 kHz between source and drain terminals as a function of the channel bias voltage, $V_{CH}$ at 300 K. The C does not show any hysteresis with a very small change of the C value with $V_{CH}$. This suggests that the hysteresis observed in the top gated configuration across the gate oxide is not related to graphene but to the graphene/$Al_2O_3$ interface. Figure 3(b) shows an abrupt increase in C when $V_{CH}$ increases beyond 2.5 V. A sudden increase in C suggests that it may be related to the topological changes in the Fermi surface due to an increase in order.[20] The asymmetry of the band structure when next-near neighbor hoping is considered can also explain this result.[21] To understand this behavior, we theoretically calculated the density of states of bilayer graphene which is shown in the inset of Fig. 3(b). The details of the calculations will be published elsewhere. The shape of C versus $V_{CH}$ qualitatively agrees with the density of states. However, the detail dependence including the anomaly above 2.5 V needs to be understood. Figure 3(c) shows $R_{xx}$ vs. $V_{TG}$ at different sweep rates at 300 K. The rate of $V_{TG}$ sweep is varied by changing the hold time between successive $V_{TG}$ increments. The hysteresis is not dependent on the sweep rate as all the curves are falling on the same curve. This indicates that interface traps are charged in time scales much smaller than the sweep speeds. Figure 3(d) shows $R_{xx}$ vs. $V_{TG}$ at different sweep rates at 300 K with an opposite sweep direction to that in Fig. 3(c). It shows that the hysteresis is not dependent on the sweep direction, neither the sweep rate. The inset of Fig. 3(d) shows $R_{xx}$ as a function of temperature (*T*) with $V_{TG} = 0$. The sample shows a metal to insulator transition at a temperature ~ 248 K which indicates that the sample may consist of electron-hole puddles, which is often reported from low mobility samples.[22] If Joule heating is the origin of the hysteresis, the direction of the hysteresis loop should be opposite below and above 248 K. The direction of the hysteresis loop in the top gated sample is the same at different temperatures as shown from Fig. 1 (c) and Fig. 2(a), indicating that Joule heating is not playing any role in inducing the hysteresis as this could have caused a change in the sequence of the hysteresis loop at 300 and 3.8 K.

In summary, capacitance - voltage measurements on top gated bilayer graphene indicates that the origin of hysteresis in the channel resistance is due to charge traps present in the graphene/$Al_2O_3$ interface with a charging and discharging time constant of ~100 μs. On the other hand, the measured capacitance of graphene between source and drain with source-drain voltage does not show any hysteresis. It is also found that the hysteresis is present even at high vacuum conditions and cryogenic temperatures indicating that chemical attachment is not the main source of the hysteresis. The hysteresis is not due to Joule heating effect, but is a function of the level of the applied voltage.

This work was partially supported by the Singapore NRF-CRP 4-2008-06.

**Figure Captions:**

FIG. 1. (a) Optical micrograph of the patterned graphene device (scale bar: 5µm). In the figure "S" stands for source, "G" for top gate, and "D" for drain. (b) Raman spectrum of bilayer graphene. The inset in (b) shows the 2D peak along with a theoretical fit. (c) Channel resistance ($R_{xx}$) vs. top gate voltage ($V_{TG}$) of the bilayer graphene at 300 K. The measurements are done at different range voltages. (d) Capacitance ($C$) vs. $V_{TG}$ at 300 K performed at 10 kHz with an AC amplitude of 500 mV.

FIG. 2. (a) Channel resistance ($R_{xx}$) vs. top gate voltage ($V_{TG}$) at 3.8 K. (b) Capacitance ($C$) vs. $V_{TG}$ at 3.8 K. The measurements in (b) are performed at 10 kHz and an AC amplitude of 500 mV. (c) $C$ vs. frequency $f$ as a function of $V_{TG}$ at 300 K. (d) $C$ vs. $V_{TG}$ at $f$ = 100 kHz and 1 MHz at 300 K.

FIG. 3. (a) Capacitance ($C$) vs. source-drain voltage ($V_{CH}$) at 300 K. (b) $C$ vs. $V_{CH}$ in the range of -3 to 3 V at 300 K. The measurements of $C$ are performed at 10 kHz and an AC amplitude of 200 mV. The inset in (b) shows the density of states (DOS) of the bilayer graphene as a function of energy. Plot of channel resistance ($R_{xx}$) vs. top gate voltage ($V_{TG}$) at 300 K at different sweep rates (dV/dt) in one sweep direction (c) and the opposite sweep direction (d) of the loop indicated by arrows. The inset in (d) shows $R_{xx}$ vs. temperature (T).

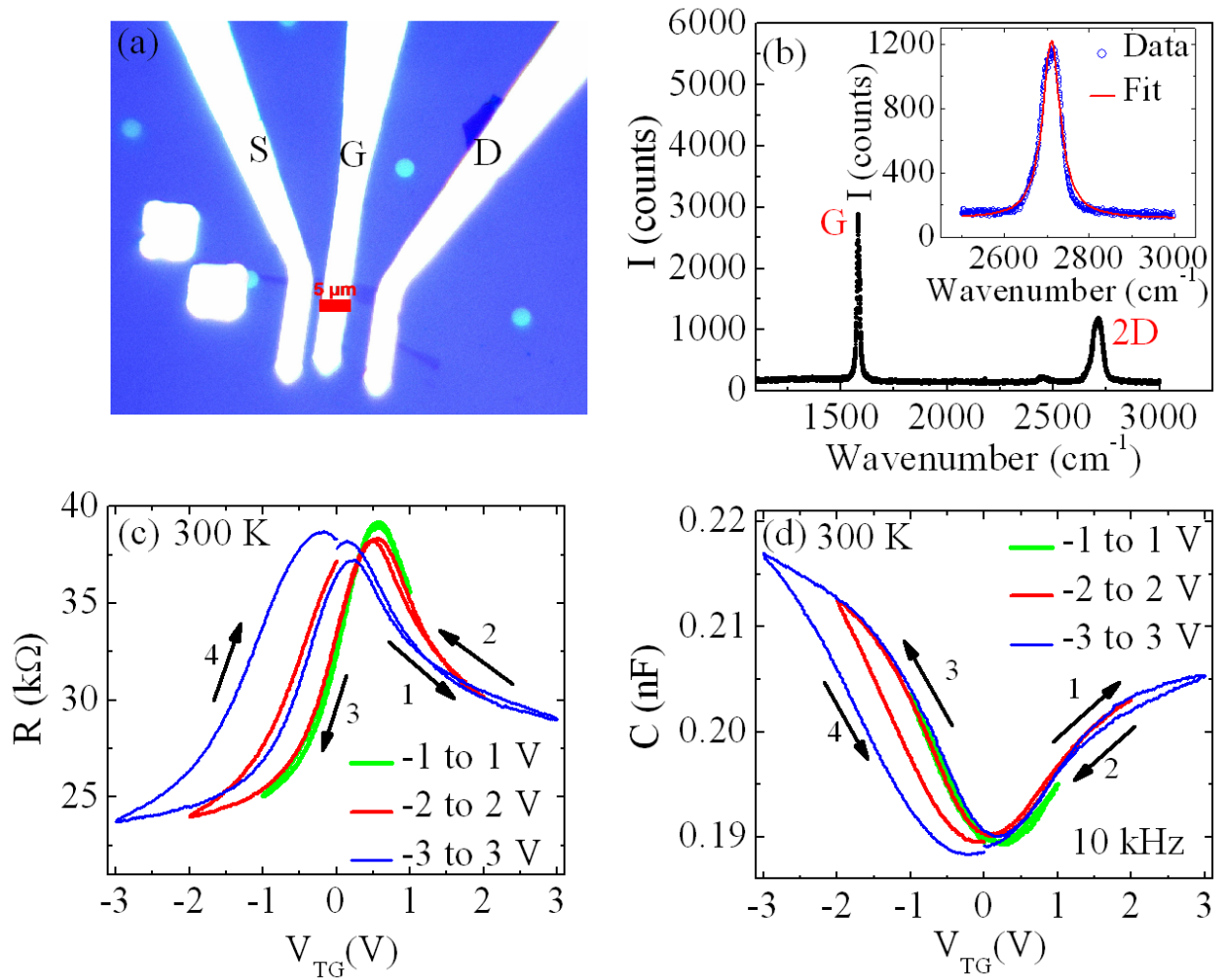

Figure 1

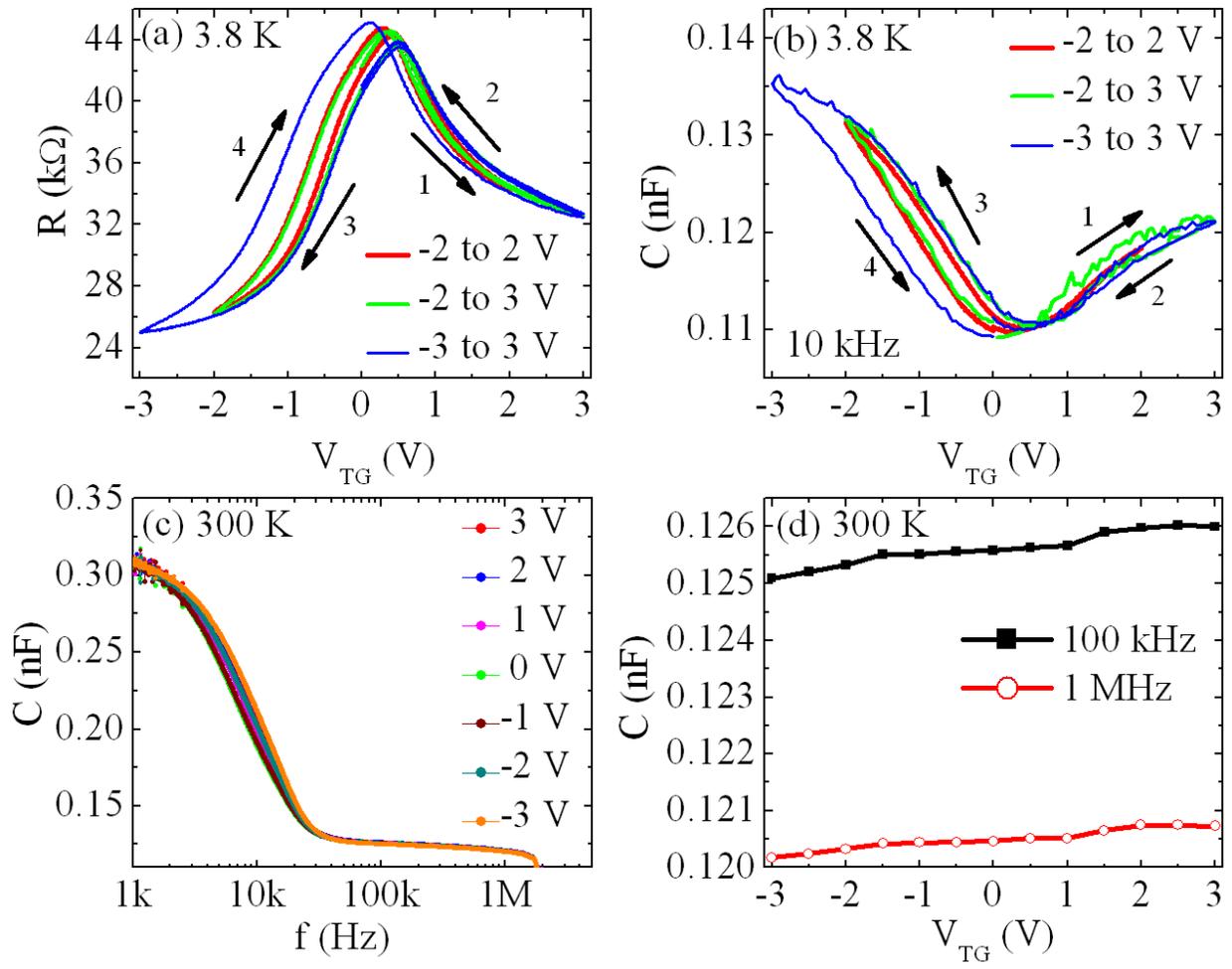

Figure 2

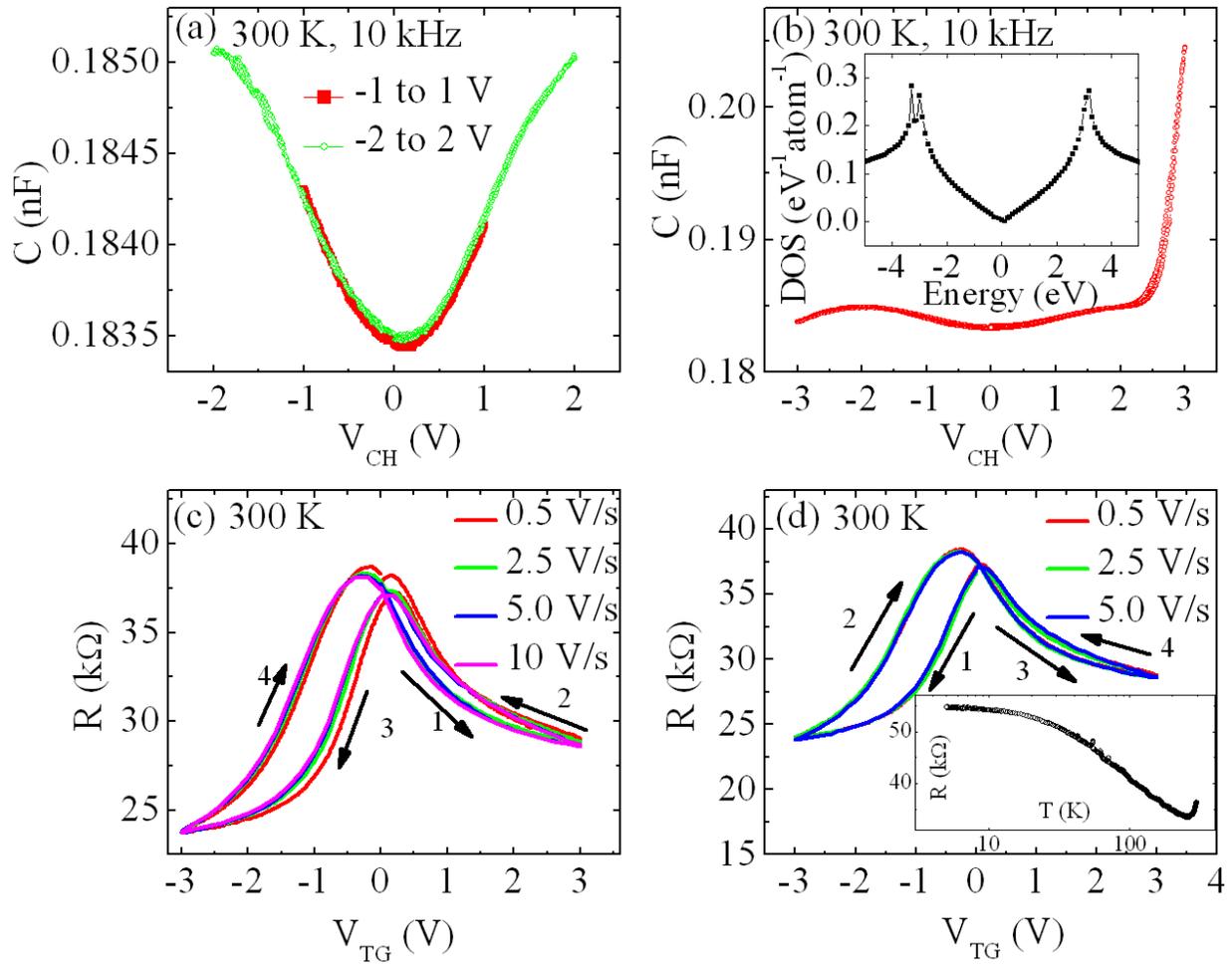

Figure 3